\documentclass[twocolumn, showpacs, preprintnumbers, amsmath, amssymb]{revtex4}

\usepackage{graphicx}
\usepackage{dcolumn}
\usepackage{bm}

\begin{document}


\title{Optical study on doped polyaniline composite films}

\author{G. Li}
\author{P. Zheng}%
\author{N. L. Wang}
\email{nlwang@aphy.iphy.ac.cn}
\author{Y. Z. Long}
\author{Z. J. Chen}
\affiliation{
Institute of Physics and Center for Condensed Matter Physics, Chinese Academy of Sciences, P. O. Box 603, Beijing, 100080, P. R. China\\
}%

\author{J. C. Li}
\author{M. X. Wan}
\affiliation{
Organic Solid Laboratory, Center for Molecular Sciences, Institute of Chemistry, Chinese Academy of Sciences, Beijing, 10080, P. R. China\\
}%

\date{\today}

\begin{abstract}
Localization driven by disorder has a strong influence on the
conducting property of conducting polymer. A class of authors hold
the opinion that disorder in the material is homogeneous and
conducting polymer is disordered metal close to Anderson-Mott
Metal-Insulator transition, while others treat the disorder as
inhomogeneous and have the conclusion that conducting polymer is a
composite of ordered metallic regions and disordered insulating
regions. The morphology of conducting polymers is an important
factor that have influence on the type and extent of disorder.
Different protonic acids used as dopants and moisture have
affection on polymer chain arrangement and interchain
interactions. A PANI-CSA film, two PANI-CSA/PANI-DBSA composite
films with different dopants ratio, and one of the composite films
with different moisture content are studied. Absolute reflectivity
measurements are performed on the films. Optical conductivity and
the real part of dielectric function are calculated by
Kramers-Kronig(KK) relations. $\sigma_1(\omega)$ and
$\varepsilon_1(\omega)$ derivate from simple Drude model in low
frequency range and tendencies of the three sample are different
and non-monotonic. The Localization Modified Drude model(LMD) in
the framework of Anderson-Mott theory can not give a good fit to
the experimental data. By introducing a distribution of relaxation
time into LMD, reasonable fits for all three samples are obtained.
This result supports the inhomogeneous picture.
\end{abstract}

\maketitle

\section{Introduction}

Doped conducting polymers are widely studied since 1980s. Although
the room temperature DC conductivity of these material has reached
that of normal metals, and some other metallic features like
finite zero temperature conductivity, negative dielectric constant
are observed, the quasi-1D nature of polymer chains indicates that
its conducting mechanism is different from conventional metals.
Because conducting polymer samples are piled up by a large number
of polymer chains, the morphology of sample has important
influence on its conductivity. Disorder is usually discussed and
remains controversial about whether it is homogeneous or
inhomogeneous in conducting polymers.\cite{epstein, heeger}

It is known that disorder is determined by several factors, mainly
including dopants used, sample preparing procedure and later
treatments. Different protonic acids used as dopants have
different molecular sizes, weights and electronegativity, and thus
will affect polymer chain arrangement and interchain interactions.
Compared with other factors, dopants can be quantitatively
controlled more easily. In this study, dopants used are camphor
sulfonic acid(CSA) and dodecylbenzene sulphonic acid(DBSA).  A
PANI-CSA film and two PANI-CSA/PANI-DBSA composite films with
different dopants ratio are studied, also one of the composite
films with different moisture content. Optical reflectivity
measurements are performed on the samples. $\rho_{DC}(T)$ provides
a basic understanding of conducting properties and are used to
identify different transport regimes in the existence of
Metal-Insulator Transition (MIT). However, macroscopic T
dependence of DC resistivity may not correspond to only one
microscopic transport mechanism, and in conducting polymers there
may be several charge transfer processes with different time
scales coexisting.\cite{synmet12543} Reflectivity spectra, along
with optical conductivity and the real part of dielectric function
calculated by KK relations, could probe the response of electron
system over a large energy range and different time scales. It is
an effective way for the investigation of charge transport
mechanism and several intensive studies on reflectivity of
conducting polymers exist. Although the reported reflectivity data
have similar features, analysis and explanation of optical
conductivity and the real part of dielectric function remain
controversial, especially in low frequency behaviors.
\cite{prb4814884, synmet84709, prb66085202, prb6013479} We
observed different and non-monotonic tendency in low frequency
range of the samples and tried to explain the behavior in a
unified framework.

\section{Experiment}
Aniline monomer is polymerized in solution in the presense of
protonic acid(CSA/DBSA) as dopant, then Ammonium persulfate as
oxidant is dissolved in deionized water and slowly added into
previously cooled mixture. After all the oxidant is added, the
reaction mixture is stirred for 24h. The precipitate is then
washed with deionized water, methanol and ethylether separately
for several times, and dried at room temperature in a dynamic
vacumm for 24h to finally obtain the powder of doped polyaniline.
To prepare porous PANI-CSA/PANI-DBSA composite films, preparation
of PANI-CSA m-cresol solution and preparation of PANI-DBSA
chloroform solution are done separately and then the two solutions
with appropriate ratio were mixed and combined with supersonic
stirring. Porous films were obtained by casting the blend solution
onto a glass plate. After drying at room temperature in air the
polyblend was peeled off the glass substrate to form a
free-standing film.

The near-normal incidence reflectance spectra were measured by
using a Bruker IFS66v/S spectrometer in the frequency range from
40$cm^{-1}$ to 25000$cm^{-1}$. The sample was mounted on an
optically black cone in a cold-finger flow cryostat. An \emph{in
situ} overcoating techinque was employed for reflectance
measurement, \cite{appop322973}which could remove the effect
caused by non flatness of sample surface. A series of  light
sources,beam splitters and detectors were used in different
frequency ranges.The connections between different regions were
excellent because of the identical overlap. The optical
conductivity and dielectric constants were calculated by KK
relation of the reflectivity data. At low frequency end,
Hagen-Rubens relation was used to extrapolate data towards zero as
in most literatures. At high frequency end of measurement,
$R(\omega)$ was extrapolated using $R\propto\omega^{-2}$ to 300000
$cm^{-1}$, and beyond that a free electron behavior of
$R\propto\omega^{-4}$ was used.

\section{Results and Discussion}
\begin{figure}
\centering
\includegraphics[width=9.5cm]{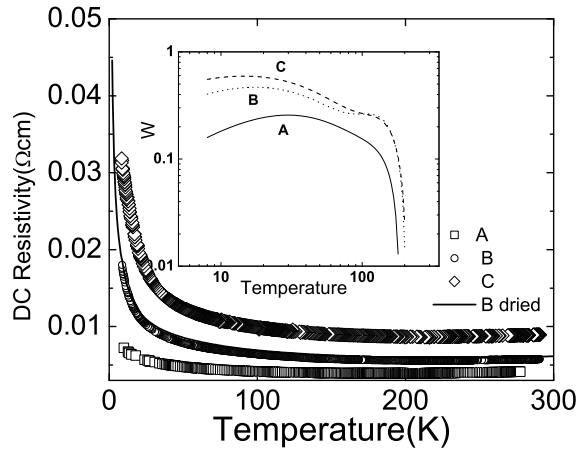}
\caption{\label{fig:resistivity}$\rho_{DC}$ of sample A, B and C,
from 8-300K. B dried is the DC resistivity of sample B measured
after storing in ambient air for several months, down to 2K. There
is a tiny elevation in value.Inset is the activation energy
$W=d(ln\sigma)/d(lnT)$ for the three samples. $\rho_{DC}(T)$ of
all samples are first fitted by six order polynomial and later
calculation is based on the fitting formula, the plot range is
8-300K}
\end{figure}

It is known from reported data that PANI-CSA is more conductive
than PANI-DBSA \cite{synmet128167, longyunze}, because of its
smaller counterion size and therefore stronger interchain
interactions. We label the pure PANI-CSA film,  the 15\verb+%+
PANI-CSA/85\verb+%+PANI-DBSA blend,
the 5\verb+%+PANI-CSA/95\verb+%+
PANI-DBSA blend sample A, B, C, respectively, in the later part of
this paper. Temperature-dependent DC conductivity measurements
show that $\rho_{DC}|_{C}>\rho_{DC}|_{B}>\rho_{DC}|_{A}$, as in
Fig.\ref{fig:resistivity}. The activation energy
$W=d(ln\sigma)/d(lnT)$ is used as a more effective criteria in the
existence of a MIT \cite{prb4817685}. The slope of the plot is
positive, negative and constant at low temperature for sample in
the metallic, the insulating and the critical regime,
respectively. Inset of Fig. \ref{fig:resistivity} is the W vs.T
plot. Sample A has a positive slope at low temperature, confirming
that there are delocalized states at the Fermi level as
$T\rightarrow 0$ and this sample is in the metallic region of MIT.
The slope of sample B and C has weak T dependence indicating they
are near the critical regime of MIT.

\begin{figure}
\centering
\includegraphics[width=9.5cm]{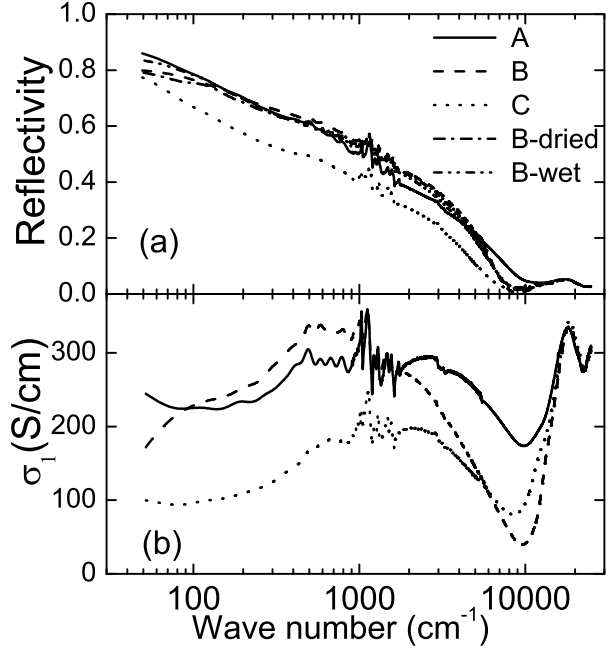}
\caption{\label{fig:randsigma} Part (a) is the reflectivity data
for all samples, part (b) is the real part of
$\sigma(\omega)=\sigma_{1}(\omega)+\imath\sigma_{2}(\omega)$ of
sample A, B and C, note the difference tendency in low frequency
range}
\end{figure}

Part (a) of Fig.\ref{fig:randsigma} is the measured reflectivity
data. At the low frequency end, the subsequence of the
reflectivity magnitude of the three samples is the same as that of
their $\rho_{DC}(Room T)$. Part (b) is the real part of
$\sigma(\omega)=\sigma_{1}(\omega)+\imath\sigma_{2}(\omega)$,
which is obtained through KK relations, with similar features of
reported data of a series of protonic acid doped PANI and Ppy
samples \cite{prb4814884, prb66085202, JPC136297, synmet68287,
prb524779, prb6013479, prb68035201}. The peak around $19000
cm^{-1}$ corresponds to $\pi-\pi^{\ast}$ interband transition.
There are a series of sharp peaks between 1000-1800 $cm^{-1}$,
which are accounted for as phonon features, the peak position
corresponding to certain bond vibration modes was given in
literature elsewhere \cite{synmet128167}. Between 1000 and 10000
$cm^{-1}$, $\sigma_{1}$ has a Drude type behavior. Below
1000$cm^{-1}$, disregarding the phonon features, $\sigma_{1}$
deviates from Drude model that it decreases with decreasing
frequency. This deviation is generally considered as the effect of
localization of electron wave functions. At the far IR region,
$\sigma_{1}(\omega)$ of the three samples have different variation
tendency. $\sigma_{1}(\omega)$ of sample A and C begin to increase
below about $100 cm^{-1}$, while $\sigma_{1}(\omega)$ of sample B
remains decreasing till the low frequency edge of our measurement.
This character is more obvious in linear axes.  Additionally,
$\sigma_{1}(\omega)|_{B}$ is larger than $\sigma_{1}(\omega)|_{A}$
in $100-1000 cm^{-1}$ range. This behavior of optical conductivity
is not consistent with $\rho_{DC}(T)$, which scales with the ratio
of dopping protonic acids used, and suggests that the charge
transport process in conducting polymer cannot be fully manifested
in DC resistivity.

\begin{figure}
\centering
\includegraphics[width=10cm]{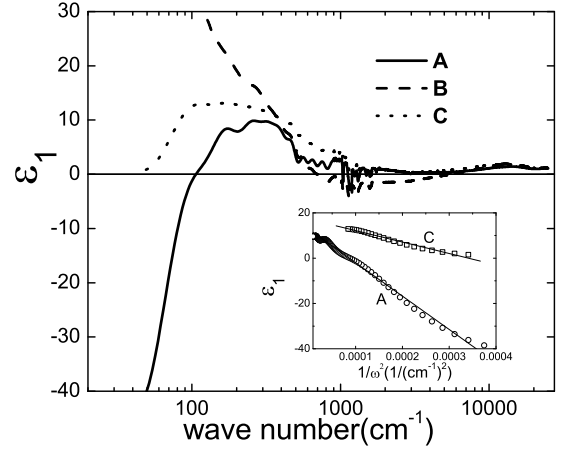}
\caption{\label{fig:epsilon} The real part of dielectric function
$\varepsilon(\omega)=\varepsilon_{1}(\omega)+\imath\varepsilon_{2}(\omega)$,
$\varepsilon_{1}$ of sample A and C have a turnover and
$\varepsilon_{1}$ of A becomes negative below 100$cm^{-1}$. Inset
is the $\varepsilon_{1}$ vs.$1/\omega^{2}$ plot for sample A and C
in low frequency range. Note that a linear relationship between
$\varepsilon_{1}$ and $1/\omega^{2}$ is Drude type response:
$\varepsilon_{\omega}=\varepsilon_{\infty}-\frac{\omega^{2}_{p1}}{\omega^{2}}$.
The fitting $\omega_{p1}$ is 384 $cm^{-1}$, 225 $cm^{-1}$ for
sample A and C, respectively.}
\end{figure}

There is large contradiction in the reported
$\varepsilon_{1}(\omega)$ of
$\varepsilon(\omega)=\varepsilon_{1}(\omega)+\imath\varepsilon_{2}(\omega)$
derived from KK. Some authors reported large negative value of
$\varepsilon_{1}(\omega)$ in the far infrared range
\cite{prl74773, prl772766, prl783915}, while others reported
positive $\varepsilon_{1}(\omega)$ in the same range
\cite{prb4814884, prb524779}. Due to suspicion to
$\varepsilon_{1}(\omega)$ near the low measurement edge obtained
through KK, all the authors had made effort to ensure the
effectiveness of data. Besides affirming the reflectivity
measurements, the direct dielectric measurements are used as
boundary condition \cite{prb492977, prb63073203, prb64201102,
prb68035201, prl90}. The consistency between $\sigma_{1}(\omega)$
and $\varepsilon_{1}(\omega)$ is also discussed, the results of KK
should stand by causality law\cite{kittel}. Fig.\ref{fig:epsilon}
is our result of $\varepsilon_{1}(\omega)$. Features in high
frequency range are similar with reported data. In far IR range,
$\varepsilon_{1}(\omega)|_{A}$ has a turnover at approximately 300
$cm^{-1}$ and a crossover at 106 $cm^{-1}$, and then becomes
negative; $\varepsilon_{1}(\omega)|_{C}$ has a turnover at about
100 $cm^{-1}$ and decrease to nearly zero at 50 $cm^{-1}$;
$\varepsilon_{1}(\omega)|_{B}$ has no turnovers and remains
increasing with decreasing frequency. Here, the low frequency
behavior shows non-monotonic change again. It is noted that
changes in $\sigma_{1}(\omega)$ is correlated with changes in
$\varepsilon_{1}(\omega)$. Increasing at low frequency in
$\sigma_{1}(\omega)$ is a Drude type behavior, corresponding to
negative $\varepsilon_{1}(\omega)$ values because the polarization
of electron systems is out of phase with the external field.
Considering the background $\varepsilon_{\infty}$of conducting
polymers, $\varepsilon_{1}(\omega)$ should decrease with
decreasing frequency at low energy and become negative after a
crossover. Inset of Fig. \ref{fig:epsilon} is the
$\varepsilon_{1}(\omega)$ vs. $1/\omega^{2}$ plot for sample A and
C in the low frequency range, confirming the existence of Drude
type behavior. We hence hold that our $\sigma_{1}(\omega)$ and
$\varepsilon_{1}(\omega)$ data are reasonable.

\begin{figure}
\centering
\includegraphics[width=9cm]{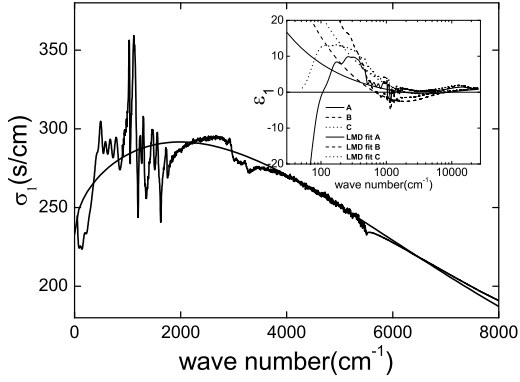}
\caption{\label{fig:lmdfit} The LMD model fit of sample A, range
from 50-8000 $cm^{-1}$. Inset is comparasion of
$\varepsilon_{1}(\omega)$ given by the LMD model using same
parameters of $\sigma_{1}(\omega)$ fit and experimental data. For
sample A and C,  the fitting curve and experimental data have
opposite behaviors in low frequency range}
\end{figure}

Hopping behavior observed in $\sigma_{DC}(T)$\cite{prb505196} and
deviation from Drude Model in $\sigma_{1}(\omega)$ and
$\varepsilon_{1}(\omega)$ suggests that localization must be
considered in conducting polymers. Because the mesoscopic
morphology of samples are  tangly built up network of polymer
chains, the main source of localization should be structural
disorder. Whether the disorder is homogenous or inhomogenous is
still in debate. Although our $\sigma_{1}(\omega)$ and
$\varepsilon_{1}(\omega)$ data have give us some hints of
inhomogeneity, we still begin our fit using homogenous model
because of calculation simplicity. In the homogenous disorder
model, Localization Modified Drude model(LMD) in the framework of
Anderson-Mott localization theory is widely used to fit
$\sigma_{1}(\omega)$ and $\varepsilon_{1}(\omega)$ \cite{heeger}:
\begin{subequations}\label{LMD}
    \begin{eqnarray}
        &&\sigma_{LMD}(\omega)=\frac{\omega_{p}^{2}\tau}{4\pi(1+\omega^{2}\tau^{2})}\nonumber \\
        &&\times[1-\frac{C}{(k_{F}v_{F})^{2}\tau^{2}}+\frac{C}{(k_{F}v_{F})^{2}\tau^{3/2}}(3\omega)^{1/2}]
    \end{eqnarray}
    \begin{eqnarray}
        &&\varepsilon_{LMD}(\omega)=\varepsilon_{\infty}+\frac{\omega_{p}^{2}\tau^2}{1+\omega^{2}\tau^{2}}\nonumber \\
        &&\times[\frac{C}{(k_{F}v_{F})^{2}\tau^{2}}(\sqrt{\frac{3}{\omega\tau}}-(\sqrt{6}-1))-1]
    \end{eqnarray}
\end{subequations}
where $\omega_{p}$ is the plasma frequency, $k_{F}$ is the Fermi
wavevector, $v_{F}$ is the Fermi velocity, $\tau$ is the
relaxation time and $\varepsilon_{\infty}$ is the high energy
dielectric constant,  C is a universal constant($\sim 1$). Fig.
\ref{fig:lmdfit} is the plot of $\sigma_{1}(\omega)|_{A}$and its
LMD fit, range from $50--1000 cm^{-1}$. Disregarding the phonon
features, the LMD model gives a good fit to the experimental data
except in low frequency range. The fitting parameters and deduced
values for samples are in Table \ref{tab:paralmdfit}.

\begin{table}
\caption{\label{tab:paralmdfit}Fitting parameters of the LMD
model, $\sigma_{1}(0)$is the calculated zero frequency optical
conductivity. Note that they are not in good accordance with room
temperature $\sigma_{DC}$, $k_{F}\lambda$ is calculated by
$k_{F}v_{F}\times\tau$}
\begin{ruledtabular}
    \begin{tabular}{ccccccc}

        \  & $\omega_{p}$ & $1/\tau$    & $\frac{C}{(k_{F}v_{F})^{2}}$ & $\sigma_{1}(0)$ & $\sigma_{DC}(RT)$ & $k_{F}\lambda$\\
        \  sample  & $(cm^{-1})$  & $(cm^{-1})$ & $(s^2)$                      & $(S/cm)$                 & $(S/cm)$\\
        \hline
        A & 12185 & 1/0. 00013 & $1. 31e^{-31}$ & 233 & 241 & 1. 91\\
        B & 7976 & 1/0. 00033 & $1. 17e^{-30}$ & 216 & 174 & 1. 62\\
        C & 6823 & 1/0. 00028 & $1. 65e^{-30}$ & 55 & 110 & 1. 16\\
        \end{tabular}
        \end{ruledtabular}
\end{table}

The order parameter $k_{F}\lambda$ for three samples are all close
to the Ioffe-Regel criterion $k_{F}\lambda\sim 1$, indicating that
these samples are close to a MIT according with the activation
energy $W=d(ln\sigma)/d(lnT)$ plot. The values of $\tau$ have a
magnitude of $10^{-15} s$, same as previous studies. However,
there are some erratic behaviors  in $\sigma_{1}(\omega)$ which
could not be satisfactorily explained.  Sample A of the largest
$\sigma_{DC}(Room T)$has the shortest relaxation time. Relaxation
time is a reflection of the extent of disorder of material.
Whether this fact accounts for the non-monotonic variation of
$\sigma_{1}(\omega)$ in far IR is not clear . The increasing
tendency of $\sigma_{1}(\omega)$ of sample A and C in the low
energy end could not be fitted by LMD using these parameters
either. Inset of Fig. \ref{fig:lmdfit} is the plot of
$\varepsilon_{1}(\omega)$ with LMD fit using the same parameters
from $\sigma_{1}(\omega)$ fit. It is clear that
$\varepsilon_{1}(\omega)$ of LMD could not simulate the decreasing
tendency at low frequency of sample A and C. In LMD model, the low
frequency $\sigma_{1}(\omega)$ is suppressed due to localization,
and $\varepsilon_{1}(\omega)$ becomes positive because disorder
scattering reduces relaxation time. Although there is good fit for
sample B in both $\sigma_{1}(\omega)$ and
$\varepsilon_{1}(\omega)$, the Drude type behavior of sample A and
C at low frequency indicates that the LMD model is not fully
applicable in present study.  Inhomogeneity in samples must be
considered.

\begin{figure*}
\centering
\includegraphics[width=7cm, height=6cm]{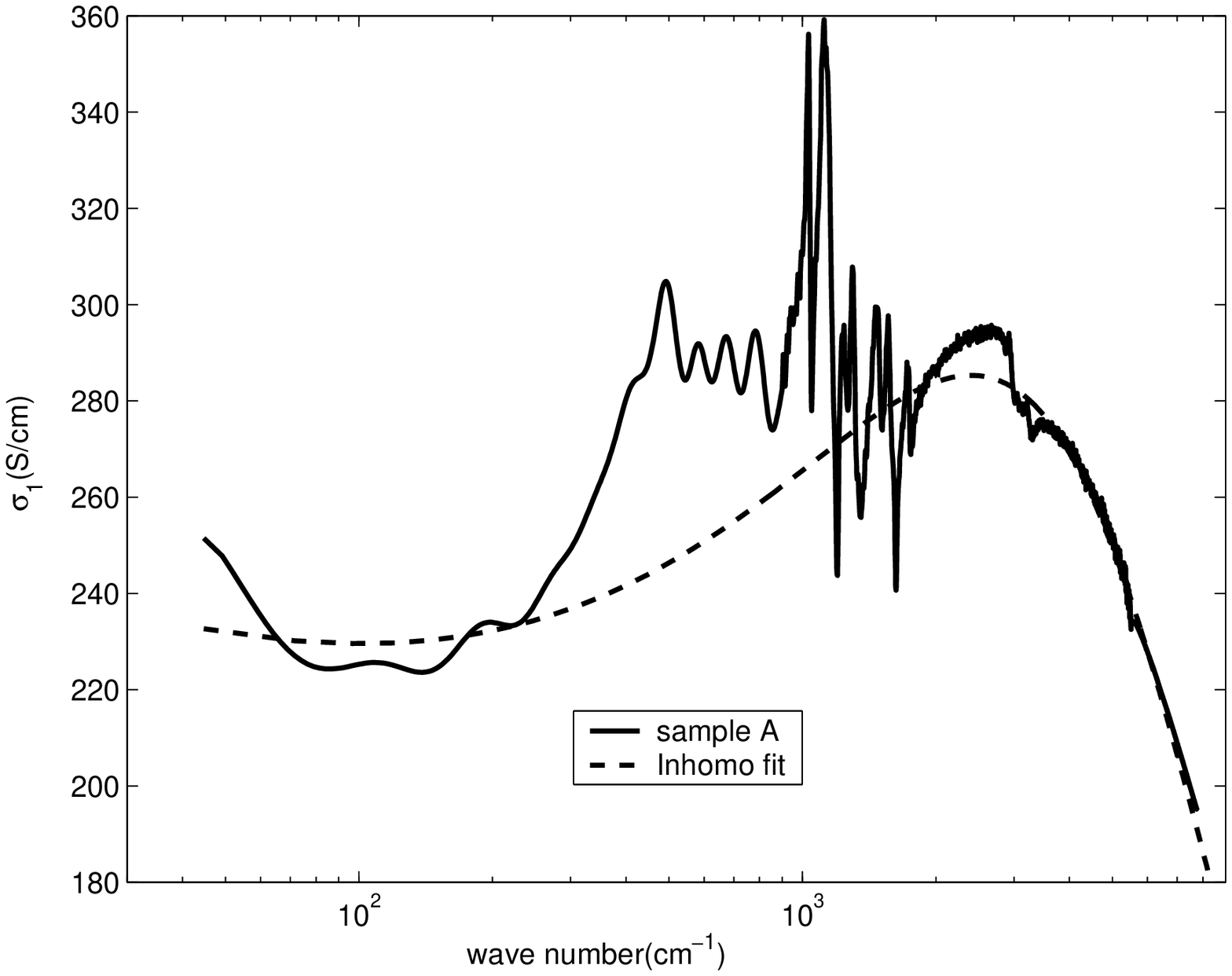}
\includegraphics[width=7cm, height=6cm]{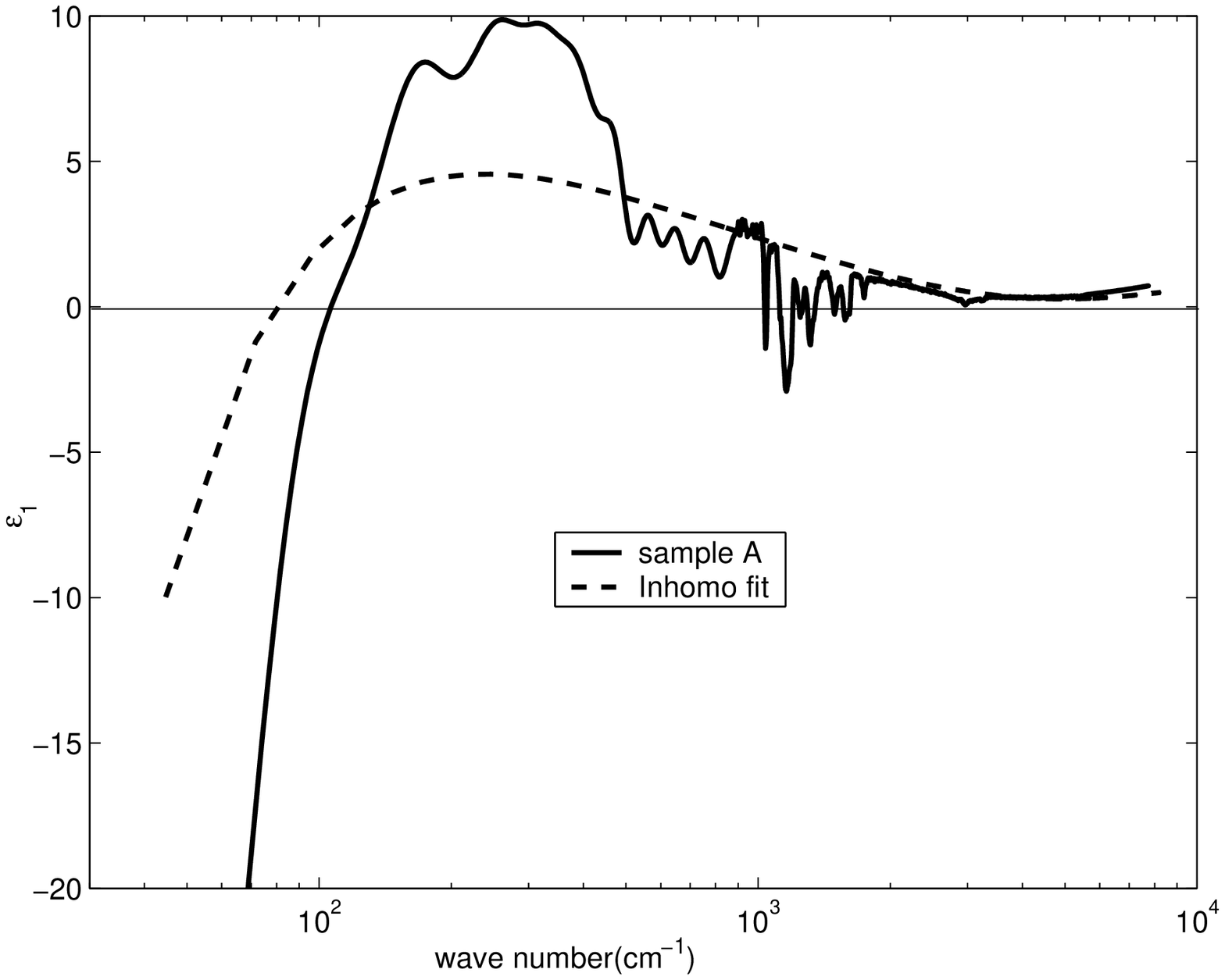}
\includegraphics[width=7cm, height=6cm]{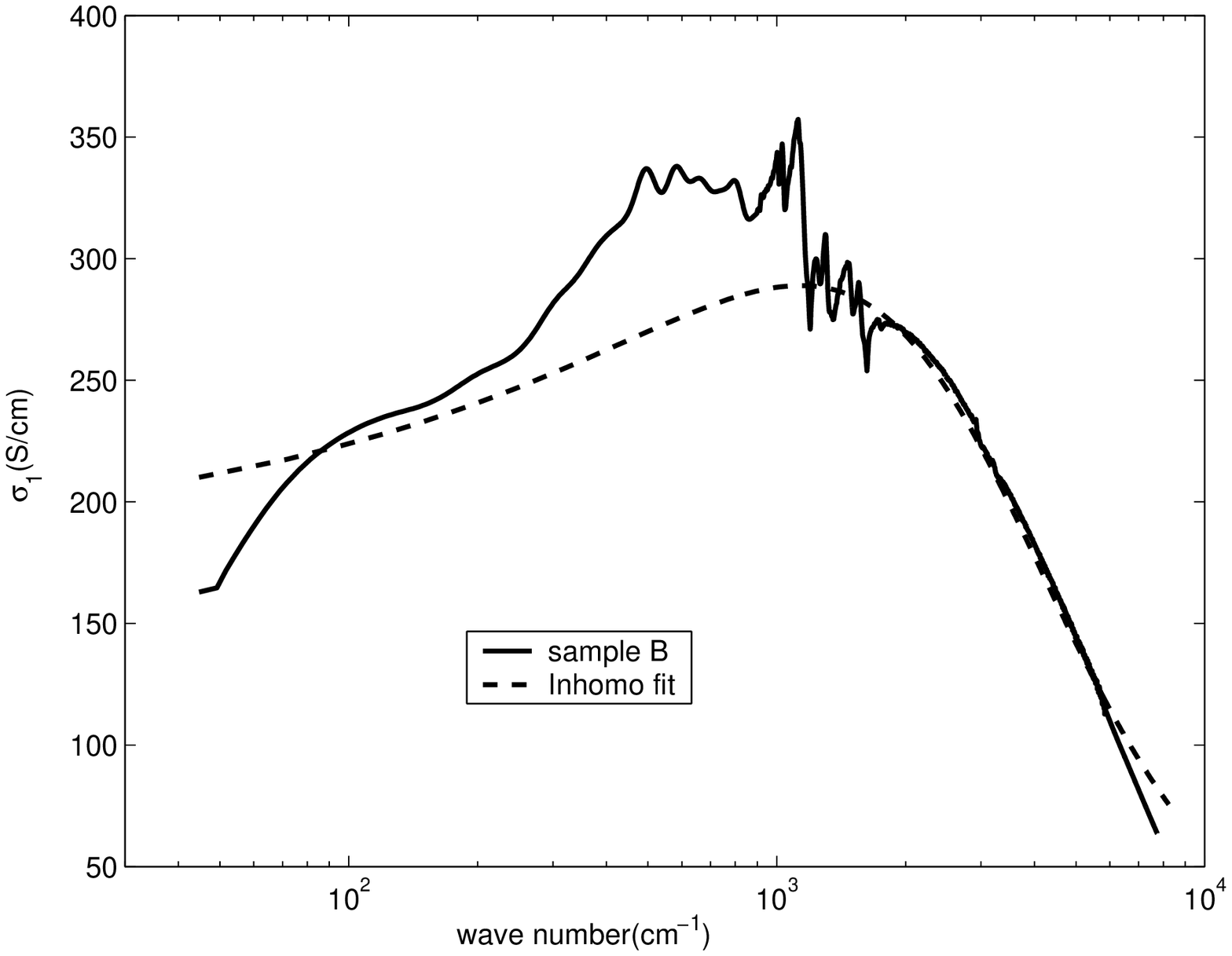}
\includegraphics[width=7cm, height=6cm]{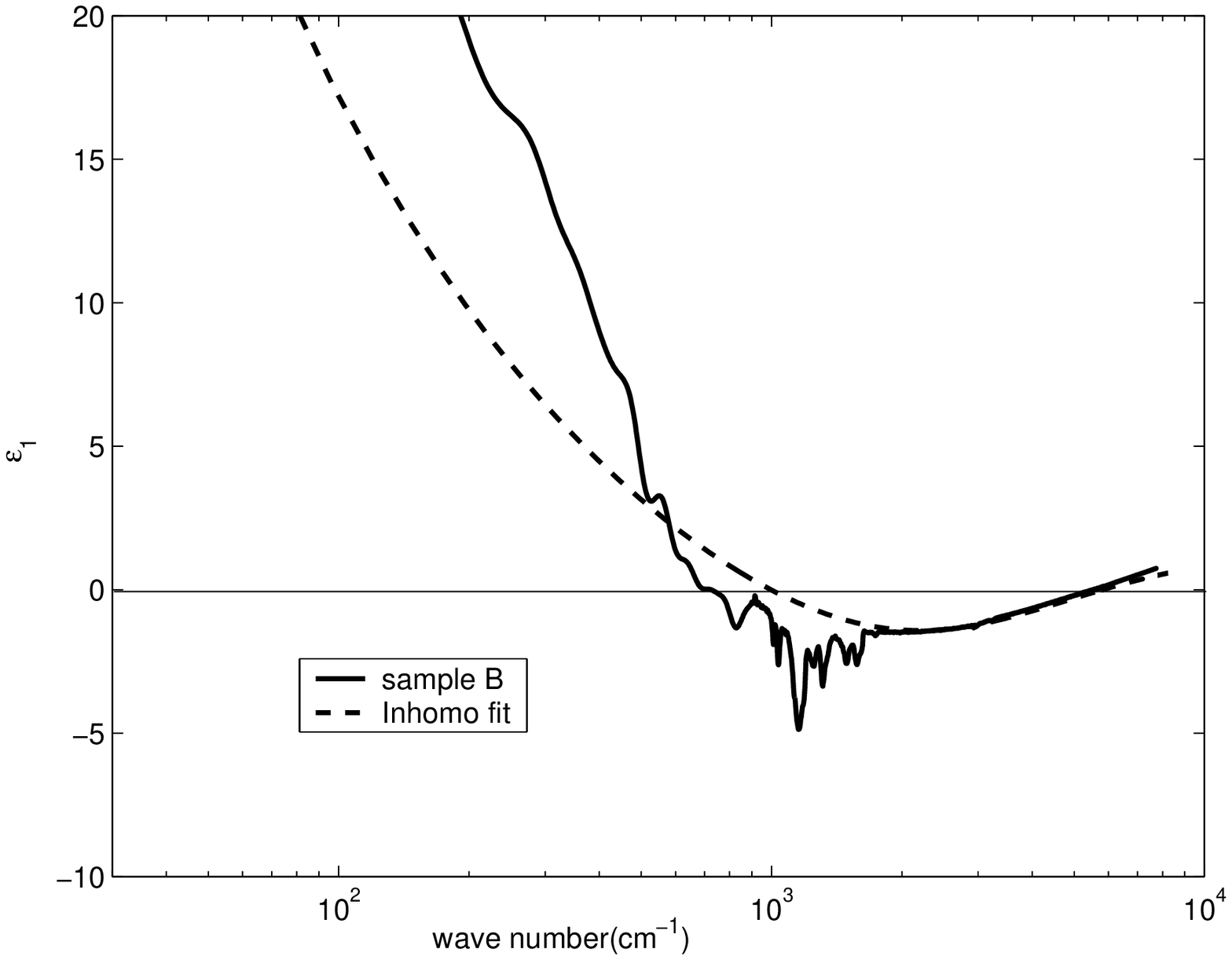}
\includegraphics[width=7cm, height=6cm]{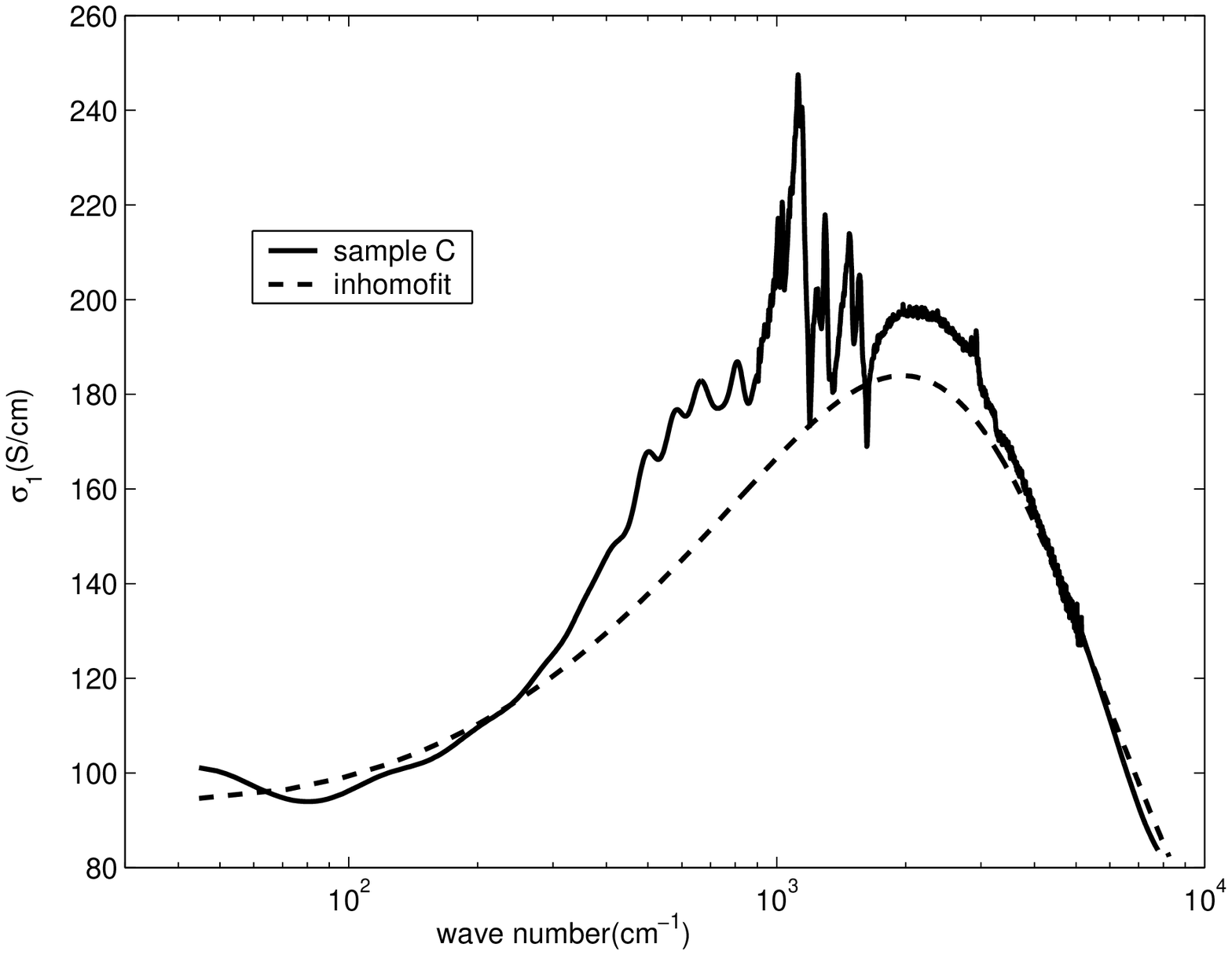}
\includegraphics[width=7cm, height=6cm]{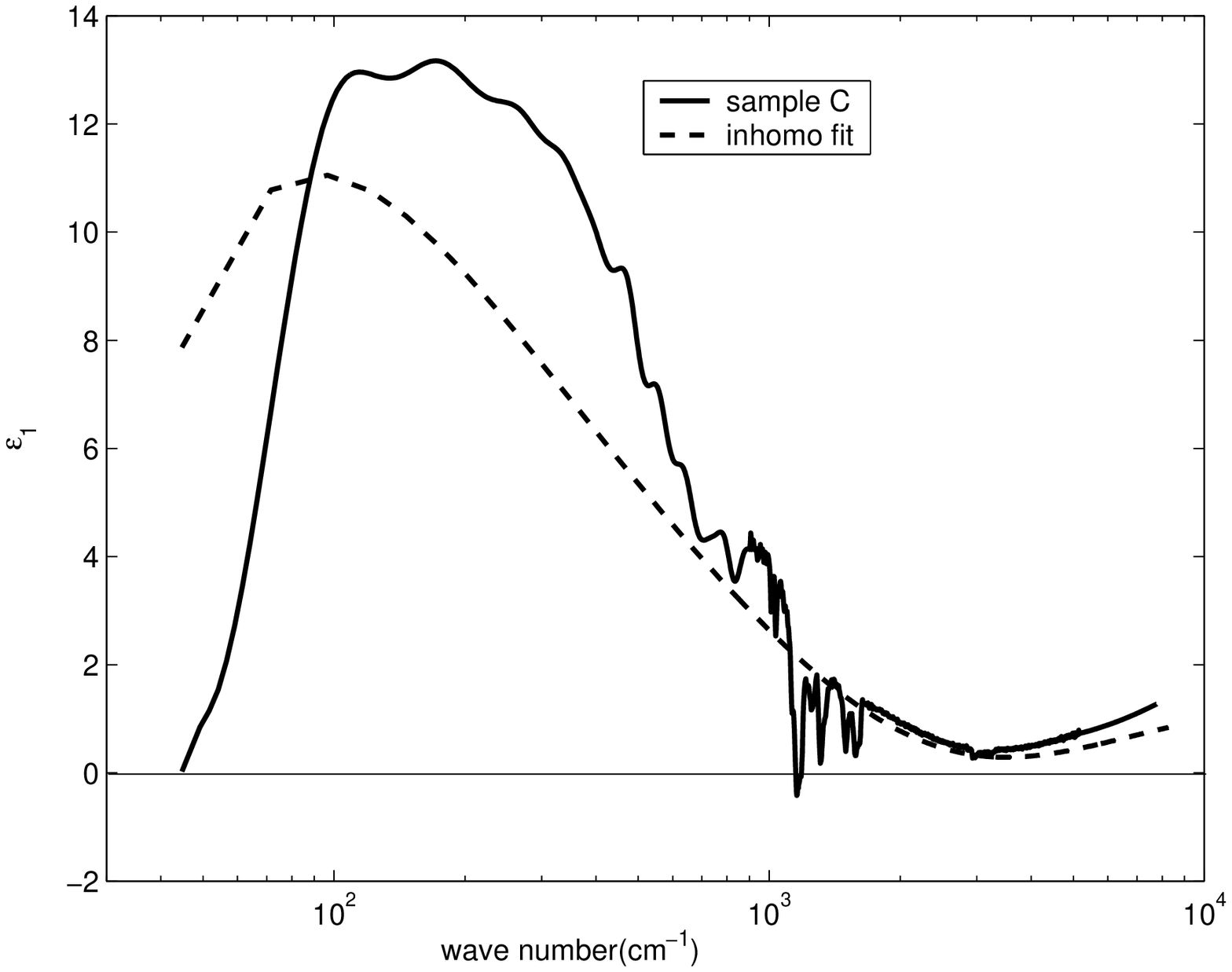}
\caption{\label{fig:inhomo}Fits of the LMD model with a
distribution of relaxation time. The Drude type behavior of sample
A and C in low frequency range are roughly simulated. There is
also a qualitatively good fit for sample B. }
\end{figure*}

In an inhomogenous picture, conducting polymers are treated as
composite materials containing mesoscopic ordered regions and
amorphous regions. In the ordered regions, polymer chains have
good spacial alignment and thus good interchain overlaping,
Electrons in these regions are delocalized and show metallic
behavior. In the disordered regions, quasi-one dimentional
localization plays the dominant role because of the quasi-one
dimentional nature of a single polymer chain. In this picture, the
Drude type response in $\sigma_{1}(\omega)$ and
$\varepsilon_{1}(\omega)$ at low far-IR range is explained by a
small fractions of delocalized charge carriers with very long
relaxation time$\sim 10^{-13}s$, as indicated by a very small
plasma frequency in $\varepsilon_{1}(\omega)$, while the most part
of carriers are localized. It is clear that movement of electrons
within a ordered region, between ordered regions and in disordered
regions have different mechanisms and characteristic time scales,
so it is difficult to describe the energy dependence of response
by uniform formula over a wide frequency range. Considering that
the general feature of $\sigma_{1}(\omega)$ at high frequency can
be fitted by both a homogeneous model or an inhomogenous
model\cite{prl90}, we followed ref.\cite{epstein} using a
distribution function of relaxation time $\tau$ to introduce
inhomogeneity into the LMD model. $\tau$ of most part of carriers
has a magnitude of $10^{-15}$ as Ioffe-Regel criterion
$k_{F}\lambda\sim 1$ allowed, while a small fraction of carriers
has a long $\tau$ as experimentally observed. $\sigma_{1}(\omega)$
and $\varepsilon_{1}(\omega)$ are given as:
\begin{subequations}\label{inhomo}
    \begin{eqnarray}
        \sigma_{inhomo}(\omega)&=&\int^{\infty}_{0}P(\tau)\sigma_{LMD}(\omega, \tau)d\tau \\
        \varepsilon_{inhomo}(\omega)&=&\int^{\infty}_{0}P(\tau)\varepsilon_{LMD}(\omega, \tau)d\tau
    \end{eqnarray}
\end{subequations}
$P(\tau$) is the distribution function of relaxation time.
\begin{equation}\label{distribution}
    P(\tau)=\frac{2\Delta}{\pi}\left[\frac{\tau^{2}}{(\tau^2-\tau_{0}^{2})^{2}+\tau^{2}\Delta^{2}}\right]
\end{equation}
$\tau_{0}$ is the average relaxation time, $\triangle$ is the
expansion of relaxation time.  Fig. \ref{fig:inhomo} is the
$\sigma_{1}(\omega)$ and $\varepsilon_{1}(\omega)$ fit for sample
A, B and C, fitting parameters are in Table \ref{tab:inhomopara}.
\begin{table}
\caption{\label{tab:inhomopara}Fitting parameters of the LMD model
with a distribution of relaxation time $\tau$, $\Delta$ is the
expansion of relaxation time}
\begin{ruledtabular}
\begin{tabular}{cccccc}

        \  & $\omega_{p}$ & $1/\tau_{0}$    & $\frac{C}{(k_{F}v_{F})^{2}}$ & $1/\Delta$ & $\sigma_{1}(0)$ \\
        \ sample & $(cm^{-1})$  & $(cm^{-1})$ & $(s^2)$                      & ($cm^{-1})$     & (S/cm)\\
        \hline
        A & 10900 & 1/0. 00015 & $4. 50e^{-31}$ & 1/0. 000017 & 328 \\
        B & 7880 & 1/0. 00031 & $1. 18e^{-30}$ & 1/0. 00000027 & 180 \\
        C & 6500 & 1/0. 00027 & $2. 17e^{-30}$ & 1/0. 000037 & 177 \\
  \end{tabular}
\end{ruledtabular}
\end{table}
Behaviors in low frequency range for the three samples are all
qualitatively simulated by including the distribution function of
$\tau$. The ratio $\triangle/\tau_{0}$ for sample A and C $\sim
0.1$, while that for sample B $\sim 0. 001$. As an estimation,
integration of $\tau$ from $0.5\times10^{2}\tau_{0}$ to
$5\times10^{2}\tau_{0}$ would give the fraction of carrier
concentration whose ralaxation time has the magnitude of
$10^{-13}s$, the results are 0.001 for sample A and C, and 0.0001
for sample B. Another estimation of the fraction,  is to compare
the small plasma frequency $\omega_{p1}$ derived from the low
frequency $\varepsilon_{\omega}$ vs. $1/\omega^{2}$ plot in
fig.\ref{fig:epsilon} and the plasma frequency $\omega_{p}$ from
LMD fit, as done in ref.\cite{prl772766}:
\begin{equation}\label{fraction}
\delta=(m^{*}_{1}/m^{*})(\omega_{p1}/\omega_{p})^{2}
\end{equation}
The results are 0.00099, 0.0011 for sample A and C, respectively.
Although $\omega_{p1}$  derived from such a small frequency range
and the assumption that $m^{*}_{1}\sim m^{*}$ are doubtful, we
still see that the two estimations give the same results,
indicating that there are more carriers with long relaxation time
existing in sample A and C, and increasing in $\triangle$ will
induce a turnover from positive $\varepsilon_{1}$ to negative.

\begin{figure}
\centering
\includegraphics[width=10cm]{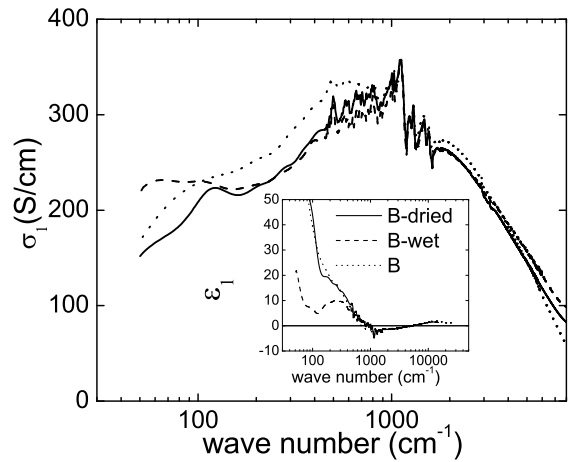}
\caption{\label{fig:moisture} Optical conductivity of sample B and
its dried and moistured form, Inset is their
$\varepsilon_{1}(\omega)$ plot}
\end{figure}

Although sample B have the largest average $\tau_{0}$, the tiny
expansion indicates that $\tau$ of most of its carriers have the
scale $10^{-15}$ as Ioffe-Regel criterion predicted, so
localization effect is dominant in sample B while sample A and C
have a fraction of carriers showing behavior of delocalized
electrons.This non-monotonicity results from the difference in the
extent of disorder, or, the intensity of intra and interchain
interactions, which is not entirely determined by the composition
of the samples. The presense of moisture has been observerd to
affect the DC conductivity of conducting polymers  significantly
\cite{synmet8917}, and the moisture influence on sample quality
should be manifested in optical data. Reflectivity measurements
are performed on two samples which have the same composition as
sample B, one is stored in ambient air for over 12 months
(B-dried), whose DC resistivity data are show in
Fig.\ref{fig:resistivity}, and the other is also stored but damped
with water just before measurements (B-wet). Fig.
\ref{fig:moisture} shows the optical conductivity for sample B,
B-dried and B-wet. $\sigma_{1}(\omega)$ of B-dried is lower than B
in far-IR range, with the similar tendency, while that of B-wet is
larger than B below 100 $cm^{-1}$. It is assumed that removal of
water molecules will reduce the structrual order between polymer
chains in the metallic regions as well as on chains bridging the
metallic regions, in an inhomogeneous picture \cite{prb5310690},
so the increasing in low frequency $\sigma_{1}(omega)$ of B-wet
can be interpreted as enhancement of tunneling between metallic
regions in low frequency \cite{synmet12543} due to reduced
potential barriers, as water molecules could reduce polarization
effects of the counter-anion, hence decrease scattering cross
section due to the counter-anions. A fit to B-wet as done in last
paragraph gives $\omega_{p}$=8500 $cm^{-1}$, $1/\tau_{0}$=0.00024
$cm^{-1}$, $c/(k_{F}v_{F})^{2}$=5.9$e^{-31}$ $s^{2}$,
$1/\Delta$=0.000008 $cm^{-1}$. The ratio
$\Delta/\tau_{0}\approx0.03$, obviously larger than that of B,
indicating that in the wet sample the concentration of carriers
with a long $\tau$ increases. However, $\sigma_{1}(\omega)$ of
B-wet is lower than B in 200-1000 $cm^{-1}$, almost the same as
B-dried, its $\varepsilon_{1}(\omega)$ (inset of
Fig.\ref{fig:moisture}) is lower than B and B-dried but is still
positive with two turnovers. This suggests that the increasing in
tunneling rate between grains can not compensate the holistic
increase of disorder, and according with the conclusion that low
frequcy behavior is determined by the competition between coherent
and incoherent channels \cite{prl90}.

\section{Conclusion}
Absolute reflectivity measurements are performed on one PANI-CSA
film, two PANI-CSA/PANI-DBSA composite films, and one of the
composite films with different moisture content. Variation of
counter-ion composition and moisture content is supposed to
modulate polymer chain arrangement and interactions. The charge
transport process in conducting polymer cannot be fully manifested
in DC resistivity. Optical conductivity and the real part of
dielectric function are calculated through Kramers-Kronig
relations, the validity of data is discussed. $\sigma_1(\omega)$
and $\varepsilon_1(\omega)$ derivate from simple Drude model in
low frequency range and tendencies of the samples are different
and non-monotonic. The localization modified Drude model in the
framework of homogeneous disorder cannot explain the behavior of
two samples. After considering the inhomogeneity by inducing a
distribution function of relaxation time into the LMD model,
$\sigma_1(\omega)$ and $\varepsilon_1(\omega)$ of the samples are
all qualitatively well fitted and explained. This result supports
the picture that disorder in the samples are inhomogeneous and the
samples are consisted of ordered metallic regions and disordered
regions.
\begin{acknowledgments}
This work is supported by National Science Foundation of China(No.
10025418), the Knowledge Innovation Project of Chinese Academy of
Sciences.
\end{acknowledgments}

\bibliography{polymer}

\begin{thebibliography}{26}
\expandafter\ifx\csname natexlab\endcsname\relax\def\natexlab#1{#1}\fi
\expandafter\ifx\csname bibnamefont\endcsname\relax
  \def\bibnamefont#1{#1}\fi
\expandafter\ifx\csname bibfnamefont\endcsname\relax
  \def\bibfnamefont#1{#1}\fi
\expandafter\ifx\csname citenamefont\endcsname\relax
  \def\citenamefont#1{#1}\fi
\expandafter\ifx\csname url\endcsname\relax
  \def\url#1{\texttt{#1}}\fi
\expandafter\ifx\csname urlprefix\endcsname\relax\def\urlprefix{URL }\fi
\providecommand{\bibinfo}[2]{#2}
\providecommand{\eprint}[2][]{\url{#2}}

\bibitem[{\citenamefont{Kohlman and Epstein}(1998)}]{epstein}
\bibinfo{author}{\bibfnamefont{R.~S.} \bibnamefont{Kohlman}} \bibnamefont{and}
  \bibinfo{author}{\bibfnamefont{A.~J.} \bibnamefont{Epstein}},
  \emph{\bibinfo{title}{Handbook of conducting polymers,2nd}},
  vol.~\bibinfo{volume}{3} (\bibinfo{publisher}{Marcel Dekker,New York},
  \bibinfo{year}{1998}).

\bibitem[{\citenamefont{Menon et~al.}(1998)\citenamefont{Menon, Yoon, Moses,
  and Heeger}}]{heeger}
\bibinfo{author}{\bibfnamefont{R.}~\bibnamefont{Menon}},
  \bibinfo{author}{\bibfnamefont{C.~O.} \bibnamefont{Yoon}},
  \bibinfo{author}{\bibfnamefont{D.}~\bibnamefont{Moses}}, \bibnamefont{and}
  \bibinfo{author}{\bibfnamefont{A.~J.} \bibnamefont{Heeger}},
  \emph{\bibinfo{title}{Handbook of conducting polymers,2nd}},
  vol.~\bibinfo{volume}{2} (\bibinfo{publisher}{Marcel Dekker,New York},
  \bibinfo{year}{1998}).

\bibitem[{\citenamefont{Prigodin and Epstein}(2002)}]{synmet12543}
\bibinfo{author}{\bibfnamefont{V.~N.} \bibnamefont{Prigodin}} \bibnamefont{and}
  \bibinfo{author}{\bibfnamefont{A.~J.} \bibnamefont{Epstein}},
  \bibinfo{journal}{Syn. Met.} \textbf{\bibinfo{volume}{125}},
  \bibinfo{pages}{43} (\bibinfo{year}{2002}).

\bibitem[{\citenamefont{Lee et~al.}(1993)\citenamefont{Lee, Heeger, and
  Cao}}]{prb4814884}
\bibinfo{author}{\bibfnamefont{K.}~\bibnamefont{Lee}},
  \bibinfo{author}{\bibfnamefont{A.~J.} \bibnamefont{Heeger}},
  \bibnamefont{and} \bibinfo{author}{\bibfnamefont{Y.}~\bibnamefont{Cao}},
  \bibinfo{journal}{Phys. Rev. B} \textbf{\bibinfo{volume}{48}},
  \bibinfo{pages}{14884} (\bibinfo{year}{1993}).

\bibitem[{\citenamefont{Kohlman
  et~al.}(1997{\natexlab{a}})\citenamefont{Kohlman, Tanner, Ihas, Min,
  MacDiarmid, and Epstein}}]{synmet84709}
\bibinfo{author}{\bibfnamefont{R.~S.} \bibnamefont{Kohlman}},
  \bibinfo{author}{\bibfnamefont{D.~B.} \bibnamefont{Tanner}},
  \bibinfo{author}{\bibfnamefont{G.~G.} \bibnamefont{Ihas}},
  \bibinfo{author}{\bibfnamefont{Y.~G.} \bibnamefont{Min}},
  \bibinfo{author}{\bibfnamefont{A.~G.} \bibnamefont{MacDiarmid}},
  \bibnamefont{and} \bibinfo{author}{\bibfnamefont{A.~J.}
  \bibnamefont{Epstein}}, \bibinfo{journal}{Syn. Met.}
  \textbf{\bibinfo{volume}{84}}, \bibinfo{pages}{709}
  (\bibinfo{year}{1997}{\natexlab{a}}).

\bibitem[{\citenamefont{Tzamalis et~al.}(2002)\citenamefont{Tzamalis, Zaidi,
  Homes, and Monkman}}]{prb66085202}
\bibinfo{author}{\bibfnamefont{G.}~\bibnamefont{Tzamalis}},
  \bibinfo{author}{\bibfnamefont{N.~A.} \bibnamefont{Zaidi}},
  \bibinfo{author}{\bibfnamefont{C.~C.} \bibnamefont{Homes}}, \bibnamefont{and}
  \bibinfo{author}{\bibfnamefont{A.~P.} \bibnamefont{Monkman}},
  \bibinfo{journal}{Phys. Rev. B} \textbf{\bibinfo{volume}{66}},
  \bibinfo{pages}{085202} (\bibinfo{year}{2002}).

\bibitem[{\citenamefont{Chapman et~al.}(1999)\citenamefont{Chapman, Buckley,
  Kemp, Kaiser, Beaglehole, and Trodahl}}]{prb6013479}
\bibinfo{author}{\bibfnamefont{B.}~\bibnamefont{Chapman}},
  \bibinfo{author}{\bibfnamefont{R.~G.} \bibnamefont{Buckley}},
  \bibinfo{author}{\bibfnamefont{N.~T.} \bibnamefont{Kemp}},
  \bibinfo{author}{\bibfnamefont{A.~B.} \bibnamefont{Kaiser}},
  \bibinfo{author}{\bibfnamefont{D.}~\bibnamefont{Beaglehole}},
  \bibnamefont{and} \bibinfo{author}{\bibfnamefont{H.~J.}
  \bibnamefont{Trodahl}}, \bibinfo{journal}{Phys. Rev. B}
  \textbf{\bibinfo{volume}{60}}, \bibinfo{pages}{13479} (\bibinfo{year}{1999}).

\bibitem[{\citenamefont{Homes et~al.}(1993)\citenamefont{Homes, Reedyk,
  Crandles, and Timusk}}]{appop322973}
\bibinfo{author}{\bibfnamefont{C.~C.} \bibnamefont{Homes}},
  \bibinfo{author}{\bibfnamefont{M.}~\bibnamefont{Reedyk}},
  \bibinfo{author}{\bibfnamefont{D.~A.} \bibnamefont{Crandles}},
  \bibnamefont{and} \bibinfo{author}{\bibfnamefont{T.}~\bibnamefont{Timusk}},
  \bibinfo{journal}{Appl.Opt.} \textbf{\bibinfo{volume}{32}},
  \bibinfo{pages}{2973} (\bibinfo{year}{1993}).

\bibitem[{\citenamefont{Lu et~al.}(2002)\citenamefont{Lu, Ng, Xu, and
  He}}]{synmet128167}
\bibinfo{author}{\bibfnamefont{X.~H.} \bibnamefont{Lu}},
  \bibinfo{author}{\bibfnamefont{H.~Y.} \bibnamefont{Ng}},
  \bibinfo{author}{\bibfnamefont{J.~W.} \bibnamefont{Xu}}, \bibnamefont{and}
  \bibinfo{author}{\bibfnamefont{C.~B.} \bibnamefont{He}},
  \bibinfo{journal}{Syn. Met.} \textbf{\bibinfo{volume}{128}},
  \bibinfo{pages}{167} (\bibinfo{year}{2002}).

\bibitem[{\citenamefont{Long et~al.}(2003)\citenamefont{Long, Chen, Wang,
  Zhang, and Wan}}]{longyunze}
\bibinfo{author}{\bibfnamefont{Y.~Z.} \bibnamefont{Long}},
  \bibinfo{author}{\bibfnamefont{Z.~J.} \bibnamefont{Chen}},
  \bibinfo{author}{\bibfnamefont{N.~L.} \bibnamefont{Wang}},
  \bibinfo{author}{\bibfnamefont{Z.~M.} \bibnamefont{Zhang}}, \bibnamefont{and}
  \bibinfo{author}{\bibfnamefont{M.~X.} \bibnamefont{Wan}},
  \bibinfo{journal}{Physica B} \textbf{\bibinfo{volume}{325}},
  \bibinfo{pages}{208} (\bibinfo{year}{2003}).

\bibitem[{\citenamefont{Menon et~al.}(1993)\citenamefont{Menon, Yoon, Moses,
  and Heeger}}]{prb4817685}
\bibinfo{author}{\bibfnamefont{R.}~\bibnamefont{Menon}},
  \bibinfo{author}{\bibfnamefont{C.~O.} \bibnamefont{Yoon}},
  \bibinfo{author}{\bibfnamefont{D.}~\bibnamefont{Moses}}, \bibnamefont{and}
  \bibinfo{author}{\bibfnamefont{A.~J.} \bibnamefont{Heeger}},
  \bibinfo{journal}{Phys. Rev. B} \textbf{\bibinfo{volume}{48}},
  \bibinfo{pages}{17685} (\bibinfo{year}{1993}).

\bibitem[{\citenamefont{Tzamalis et~al.}(2001)\citenamefont{Tzamalis, Zaidi,
  Homes, and Monkman}}]{JPC136297}
\bibinfo{author}{\bibfnamefont{G.}~\bibnamefont{Tzamalis}},
  \bibinfo{author}{\bibfnamefont{N.~A.} \bibnamefont{Zaidi}},
  \bibinfo{author}{\bibfnamefont{C.~C.} \bibnamefont{Homes}}, \bibnamefont{and}
  \bibinfo{author}{\bibfnamefont{A.~P.} \bibnamefont{Monkman}},
  \bibinfo{journal}{J.Phys:Condens.Matter} \textbf{\bibinfo{volume}{13}},
  \bibinfo{pages}{6297} (\bibinfo{year}{2001}).

\bibitem[{\citenamefont{Lee et~al.}(1995{\natexlab{a}})\citenamefont{Lee,
  Reghu, Yuh, Sariciftci, and Heeger}}]{synmet68287}
\bibinfo{author}{\bibfnamefont{K.}~\bibnamefont{Lee}},
  \bibinfo{author}{\bibfnamefont{M.}~\bibnamefont{Reghu}},
  \bibinfo{author}{\bibfnamefont{E.~L.} \bibnamefont{Yuh}},
  \bibinfo{author}{\bibfnamefont{N.~S.} \bibnamefont{Sariciftci}},
  \bibnamefont{and} \bibinfo{author}{\bibfnamefont{A.~J.}
  \bibnamefont{Heeger}}, \bibinfo{journal}{Syn. Met.}
  \textbf{\bibinfo{volume}{68}}, \bibinfo{pages}{287}
  (\bibinfo{year}{1995}{\natexlab{a}}).

\bibitem[{\citenamefont{Lee et~al.}(1995{\natexlab{b}})\citenamefont{Lee,
  Menon, Yoon, and Heeger}}]{prb524779}
\bibinfo{author}{\bibfnamefont{K.}~\bibnamefont{Lee}},
  \bibinfo{author}{\bibfnamefont{R.}~\bibnamefont{Menon}},
  \bibinfo{author}{\bibfnamefont{C.~O.} \bibnamefont{Yoon}}, \bibnamefont{and}
  \bibinfo{author}{\bibfnamefont{A.~J.} \bibnamefont{Heeger}},
  \bibinfo{journal}{Phys. Rev. B} \textbf{\bibinfo{volume}{52}},
  \bibinfo{pages}{4779} (\bibinfo{year}{1995}{\natexlab{b}}).

\bibitem[{\citenamefont{Lee and Heeger}(2003)}]{prb68035201}
\bibinfo{author}{\bibfnamefont{K.}~\bibnamefont{Lee}} \bibnamefont{and}
  \bibinfo{author}{\bibfnamefont{A.~J.} \bibnamefont{Heeger}},
  \bibinfo{journal}{Phys. Rev. B} \textbf{\bibinfo{volume}{68}},
  \bibinfo{pages}{035201} (\bibinfo{year}{2003}).

\bibitem[{\citenamefont{Kohlman et~al.}(1995)\citenamefont{Kohlman, Joo, Wang,
  Pouget, Kaneko, Ishiguro, and Epstein}}]{prl74773}
\bibinfo{author}{\bibfnamefont{R.~S.} \bibnamefont{Kohlman}},
  \bibinfo{author}{\bibfnamefont{J.}~\bibnamefont{Joo}},
  \bibinfo{author}{\bibfnamefont{Y.~Z.} \bibnamefont{Wang}},
  \bibinfo{author}{\bibfnamefont{J.~P.} \bibnamefont{Pouget}},
  \bibinfo{author}{\bibfnamefont{H.}~\bibnamefont{Kaneko}},
  \bibinfo{author}{\bibfnamefont{T.}~\bibnamefont{Ishiguro}}, \bibnamefont{and}
  \bibinfo{author}{\bibfnamefont{A.~J.} \bibnamefont{Epstein}},
  \bibinfo{journal}{Phys. Rev. Lett.} \textbf{\bibinfo{volume}{74}},
  \bibinfo{pages}{773} (\bibinfo{year}{1995}).

\bibitem[{\citenamefont{Kohlman et~al.}(1996)\citenamefont{Kohlman, Joo, Min,
  MacDiarmid, and Epstein}}]{prl772766}
\bibinfo{author}{\bibfnamefont{R.~S.} \bibnamefont{Kohlman}},
  \bibinfo{author}{\bibfnamefont{J.}~\bibnamefont{Joo}},
  \bibinfo{author}{\bibfnamefont{Y.~G.} \bibnamefont{Min}},
  \bibinfo{author}{\bibfnamefont{A.~G.} \bibnamefont{MacDiarmid}},
  \bibnamefont{and} \bibinfo{author}{\bibfnamefont{A.~J.}
  \bibnamefont{Epstein}}, \bibinfo{journal}{Phys. Rev. Lett.}
  \textbf{\bibinfo{volume}{77}}, \bibinfo{pages}{2766} (\bibinfo{year}{1996}).

\bibitem[{\citenamefont{Kohlman
  et~al.}(1997{\natexlab{b}})\citenamefont{Kohlman, Zibold, Tanner, Ihas,
  Ishiguro, Min, MacDiarmid, and Epstein}}]{prl783915}
\bibinfo{author}{\bibfnamefont{R.~S.} \bibnamefont{Kohlman}},
  \bibinfo{author}{\bibfnamefont{A.}~\bibnamefont{Zibold}},
  \bibinfo{author}{\bibfnamefont{D.~B.} \bibnamefont{Tanner}},
  \bibinfo{author}{\bibfnamefont{G.~G.} \bibnamefont{Ihas}},
  \bibinfo{author}{\bibfnamefont{T.}~\bibnamefont{Ishiguro}},
  \bibinfo{author}{\bibfnamefont{Y.~G.} \bibnamefont{Min}},
  \bibinfo{author}{\bibfnamefont{A.~G.} \bibnamefont{MacDiarmid}},
  \bibnamefont{and} \bibinfo{author}{\bibfnamefont{A.~J.}
  \bibnamefont{Epstein}}, \bibinfo{journal}{Phys. Rev. Lett.}
  \textbf{\bibinfo{volume}{78}}, \bibinfo{pages}{3915}
  (\bibinfo{year}{1997}{\natexlab{b}}).

\bibitem[{\citenamefont{Joo et~al.}(1994)\citenamefont{Joo, Oblakowaki, Du,
  Pouget, Oh, Wiesinger, Min, MacDiarmid, and Epstein}}]{prb492977}
\bibinfo{author}{\bibfnamefont{J.}~\bibnamefont{Joo}},
  \bibinfo{author}{\bibfnamefont{Z.}~\bibnamefont{Oblakowaki}},
  \bibinfo{author}{\bibfnamefont{G.}~\bibnamefont{Du}},
  \bibinfo{author}{\bibfnamefont{J.~P.} \bibnamefont{Pouget}},
  \bibinfo{author}{\bibfnamefont{E.~J.} \bibnamefont{Oh}},
  \bibinfo{author}{\bibfnamefont{J.~M.} \bibnamefont{Wiesinger}},
  \bibinfo{author}{\bibfnamefont{Y.}~\bibnamefont{Min}},
  \bibinfo{author}{\bibfnamefont{A.~G.} \bibnamefont{MacDiarmid}},
  \bibnamefont{and} \bibinfo{author}{\bibfnamefont{A.~J.}
  \bibnamefont{Epstein}}, \bibinfo{journal}{Phys . Rev. B}
  \textbf{\bibinfo{volume}{49}}, \bibinfo{pages}{2977} (\bibinfo{year}{1994}).

\bibitem[{\citenamefont{Martens
  et~al.}(2001{\natexlab{a}})\citenamefont{Martens, Reedijk, Brom, de~Leeuw,
  and Menon}}]{prb63073203}
\bibinfo{author}{\bibfnamefont{H.~C.~F.} \bibnamefont{Martens}},
  \bibinfo{author}{\bibfnamefont{J.~A.} \bibnamefont{Reedijk}},
  \bibinfo{author}{\bibfnamefont{H.~B.} \bibnamefont{Brom}},
  \bibinfo{author}{\bibfnamefont{D.~M.} \bibnamefont{de~Leeuw}},
  \bibnamefont{and} \bibinfo{author}{\bibfnamefont{R.}~\bibnamefont{Menon}},
  \bibinfo{journal}{Phys. Rev. B} \textbf{\bibinfo{volume}{63}},
  \bibinfo{pages}{073203} (\bibinfo{year}{2001}{\natexlab{a}}).

\bibitem[{\citenamefont{Martens
  et~al.}(2001{\natexlab{b}})\citenamefont{Martens, Brom, and
  Menon}}]{prb64201102}
\bibinfo{author}{\bibfnamefont{H.~C.~F.} \bibnamefont{Martens}},
  \bibinfo{author}{\bibfnamefont{H.~B.} \bibnamefont{Brom}}, \bibnamefont{and}
  \bibinfo{author}{\bibfnamefont{R.}~\bibnamefont{Menon}},
  \bibinfo{journal}{Phys. Rev. B} \textbf{\bibinfo{volume}{64}},
  \bibinfo{pages}{201102} (\bibinfo{year}{2001}{\natexlab{b}}).

\bibitem[{\citenamefont{Romijin et~al.}(2003)\citenamefont{Romijin, Hupkes,
  Martens, Brom, Mukherjee, and Menon}}]{prl90}
\bibinfo{author}{\bibfnamefont{I.~G.} \bibnamefont{Romijin}},
  \bibinfo{author}{\bibfnamefont{H.~J.} \bibnamefont{Hupkes}},
  \bibinfo{author}{\bibfnamefont{H.~C.~F.} \bibnamefont{Martens}},
  \bibinfo{author}{\bibfnamefont{H.~B.} \bibnamefont{Brom}},
  \bibinfo{author}{\bibfnamefont{A.~K.} \bibnamefont{Mukherjee}},
  \bibnamefont{and} \bibinfo{author}{\bibfnamefont{R.}~\bibnamefont{Menon}},
  \bibinfo{journal}{Phys. Rev. Lett.} \textbf{\bibinfo{volume}{90}},
  \bibinfo{pages}{176602} (\bibinfo{year}{2003}).

\bibitem[{\citenamefont{Kittel}(1976)}]{kittel}
\bibinfo{author}{\bibfnamefont{C.}~\bibnamefont{Kittel}},
  \emph{\bibinfo{title}{Introduction to Solid State Physics}}
  (\bibinfo{publisher}{John Wiley, Sons}, \bibinfo{year}{1976}).

\bibitem[{\citenamefont{Zuppiroli et~al.}(1994)\citenamefont{Zuppiroli, Bussac,
  Paschen, Chauvet, and Forro}}]{prb505196}
\bibinfo{author}{\bibfnamefont{L.}~\bibnamefont{Zuppiroli}},
  \bibinfo{author}{\bibfnamefont{M.~N.} \bibnamefont{Bussac}},
  \bibinfo{author}{\bibfnamefont{S.}~\bibnamefont{Paschen}},
  \bibinfo{author}{\bibfnamefont{O.}~\bibnamefont{Chauvet}}, \bibnamefont{and}
  \bibinfo{author}{\bibfnamefont{L.}~\bibnamefont{Forro}},
  \bibinfo{journal}{Phys . Rev. B} \textbf{\bibinfo{volume}{50}},
  \bibinfo{pages}{5196} (\bibinfo{year}{1994}).

\bibitem[{\citenamefont{Kahol et~al.}(1997)\citenamefont{Kahol, Dyakonov, and
  McCormick}}]{synmet8917}
\bibinfo{author}{\bibfnamefont{P.~K.} \bibnamefont{Kahol}},
  \bibinfo{author}{\bibfnamefont{A.~J.} \bibnamefont{Dyakonov}},
  \bibnamefont{and} \bibinfo{author}{\bibfnamefont{B.~J.}
  \bibnamefont{McCormick}}, \bibinfo{journal}{Syn. Met.}
  \textbf{\bibinfo{volume}{89}}, \bibinfo{pages}{17} (\bibinfo{year}{1997}).

\bibitem[{\citenamefont{Pinto et~al.}(1996)\citenamefont{Pinto, Shah, Kahol,
  and McCormick}}]{prb5310690}
\bibinfo{author}{\bibfnamefont{N.~J.} \bibnamefont{Pinto}},
  \bibinfo{author}{\bibfnamefont{P.~D.} \bibnamefont{Shah}},
  \bibinfo{author}{\bibfnamefont{P.~K.} \bibnamefont{Kahol}}, \bibnamefont{and}
  \bibinfo{author}{\bibfnamefont{B.~J.} \bibnamefont{McCormick}},
  \bibinfo{journal}{Phys. Rev. B} \textbf{\bibinfo{volume}{53}},
  \bibinfo{pages}{10690} (\bibinfo{year}{1996}).

\end{thebibliography}

\end{document}